\documentstyle[12pt]{article}
\begin{document}

\hoffset -1.5cm\voffset-3cm\hsize 17cm\textheight 25cm 
\parindent 0ex
\def\natural{\hbox{N}\hskip-.85em \hbox{I}}
\def\naturalzero{\hbox{${\rm N}_0$}\hskip-1.05em \hbox{I}\hskip 1.05em\null}
\def\ratio{\hbox{Q}\hskip-.7em \hbox{$($}}
\def\real{\hbox{R}\hskip-.85em\hbox{I}}
\def\complex{\hbox{C}\llap{\hbox{$($}}}
                %
\def\abs#1{\vert #1\vert}
\def\kb{\rm k} 
\def\mb{\mu_B} 
\def\be{\begin{eqnarray}}
\def\ee{\end{eqnarray}}

\def\sgn#1{\hbox{sign}\left(#1\right)}

\title{\bf Quasiparticle energy spectrum in ferromagnetic
 Josephson weak links}

\author{ L. Dobrosavljevi\' c-Gruji\' c,$^{1}$ R. Ziki\'c,$^1$  and
Z. Radovi\' c$^2$}

\maketitle

{$^1$ Institute of Physics, P. O. Box 57, 11080, Belgrade, Yugoslavia}\\
{$^2$  Department of Physics, University of Belgrade, P.O. Box 368,
11001 Belgrade, Yugoslavia}\\

\begin{abstract}
The quasiparticles energy spectrum in clean ferromagnetic weak links between
conventional superconductors is
calculated.  Large peaks in DOS, due to a special case of Andreev
 reflection at the ferromagnetic barrier, correspond
to spin-splitt  bound states. Their
energies  are  obtained as a function
of the barrier thickness, exchange field strength, and of the macroscopic phase
difference $\phi$ at the link, related to the Josephson current. In the
 ground state, $\phi$ can be $0$ or $\pi$,
depending on  the ferromagnetic barrier influence. Conditions for the
appearence of the zero-energy bound states (ZES) and for the spin polarized
 ground state (SPGS) are obtained analytically. It is shown that ZES
 appear only  outside
 the weak link ground state.
\end{abstract}


{\bf Corresponding author:\\
L. Dobrosavljevi\' c-Gruji\' c,
Institute of Physics, P. O. Box 57, 11080, Belgrade, Serbia;\\
e-mail: dobrosav@phy.bg.ac.yu; zikic@ami-net.com\\
phone: +381 11 3162758; fax: +381 3162190}

\eject

\section{Introduction}

Recently, the study of  $\pi$-junctions with an intrinsic phase
difference $\phi=\pi$ in the ground state has gained much interest. Many
experiments on high-T$_c$ superconducting weak links have been
interpreted in terms of $\pi$-junctions proposed for the superconductors
with d-wave pairing \cite{grenoble}.
For conventional superconductors with $s-$wave pairing, $\pi$-junctions have
been suggested by Bulaevskii et al. \cite{5t} to arise in presence of
magnetic impurities in the barrier, in connection with spin flip
assisted coherent tunneling. Also, $\pi-$coupling was predicted
theoretically to exist in superconductor-ferromagnet ($S/F$) multilayers
by Radovi\' c et al. \cite{zoran}. The evidence for $\pi$- coupling,
manifested {\it via} characteristic oscillations of the superconducting
critical temperature $T_c$ with magnetic layer thickness,  was sought
experimentally in several
superconductor-ferromagnet \cite{strunk,drag,obi} and superconductor-spin
glass systems \cite{mer}.

Other manifestations of nontrivial  coupling ($\phi\neq0$)  could be
found in  the density of states (DOS) and, in particular, in the
appearence of the Andreev bound states at the Fermi level. Similarly as
in high-$T_c$ superconductors, these zero-energy states  (ZES) should be
manifested as the zero-bias conductance peaks (ZBCP) in tunneling
spectroscopy measurements \cite{8}. Recently, Andreev reflection has
been studied theoretically in $S/F$ tunnel junctions by de Jong and
Beenakker \cite{9}, in $S/F/S$ junctions by Kuplevakhskii and Fal'ko
\cite{10} and  by Tanaka and Kashiwaya \cite{11}, and in atomic-scale
$S/F$ superlattices by Proki\' c and al. \cite{vesna}. Recent
experiments \cite{soulen,upad} have shown that Andreev reflection can be
strongly suppressed by spin polarization in the ferromagnet. This is the
case  for point contacts with transparent barriers, whereas the spin
polarization can enhance the Andreev reflection in the $S/F/S$ junctions
with sizeable interface scattering \cite{bourgeois}.

In the previous papers we calculated DOS and spontaneous currents in
Josephson $S/F/S$ point contacts \cite{zikic,spie96}, considering the
combined effect of $d-$ wave pairing in $S$ electrodes and of the
exchange field in $F$ barrier.

In the present paper we calculate the quasiparticle density of states in
conventional superconductor  Josephson weak links with a  transparent
ferromagnetic metal barrier. The results are simpler than in the general
anisotropic  case \cite{zikic}, due to the isotropy of the order
parameter in $S$. The   phase difference $\phi$ at the link, related to
the flow of the supercurrent, in presence of the exchange energy $h$ in
$F$ acquires an additional, "magnetic" contribution. We study the change
in the quasiparticle energy spectrum induced by $\phi$ and $h$ and
discuss the appearence of  Z{E}S and SPGS.

The paper is organized as follows: in Section II we present briefly the
results of the quasiclassical theory of superconductivity for $S/F/S$
weak links:  the quasiparticles Green's functions in the barrier, the
resulting densities of states and the Josephson supercurrent. We show
how the local densities of states   can be expressed in terms of
energies of bound states and  give the equations for the bound states,
from which  follow the conditions for  Z{E}S and SPGS. We calculate
the supercurrent through the link and discuss its relationship with
Andreev bound states. Section III contains a discussion of the numerical
results, including the comparison with  the   case of the normal metal
($N$) barrier, $h=0$, and a brief conclusion.

\section{Quasiclassical Theory}

An efficient method for calculating local spectral properties of
superconductors is the quasiclassical theory of superconductivity
\cite{eil}. In the clean limit, and assuming that the magnetic influence
 on superconductivity is limited to that of the exchange
energy in  the barrier, one can use the Eilenberger quasiclassical
  equations in the
presence of the exchange field \cite{bbp,spie96,alex}.

\subsection{Quasiparticles Green's function}

In the previous papers  we have solved  the quasiclassical equations for
an $S/F/S$ weak link, with a transparent, thin and short $F$ barrier of
thickness $2d$, consisting of a monodomain ferromagnetic metal with
constant exchange energy $h$. Here we present the results for the
conventional case of $s-$wave pairing in $S$ electrodes. Both $S$ and
$F$ metals  are assumed clean, with same dispersion relations and with
same Fermi velocity ${\bf v_0}$ (electron scattering on impurities in
$S$  can be neglected if  $l\gg \xi_0$, where $l$ is the electron mean
free path and $\xi _0$ superconducting coherence length, and if $h\gg
\hbar v_0/l$ in $F$).

Taking the $x$-axis perpendicular to the barrier (Fig. 1), we assume a
step-function variation of the pair potential

\be
\Delta(x) = \Delta\Theta(-d-x) +\Delta\Theta(x-d).  \label{(2.5)}
              \ee
where $\Delta=\Delta(T)$.
In principle, the pair potential should be determined
self-consistently.
However, this would greatly complicate the calculations without shedding
much light on the exchange-interaction-induced states in the $S/F/S$
case \cite{10,dwa}, which is the main subject of this paper. In the
$S/N/S$ case, the validity of our model requires a weak proximity effect
in $S$, as it may be the case at low $T$ and for thin and short
barriers.

The quasiclassical Green's function in the barrier is
\be
g=g_{\downarrow}={{\omega_n\cos\gamma_n+i \Omega_n\sin\gamma_n}\over
{ \Omega_n\cos\gamma_n+i \omega_n\sin\gamma_n}}
\label{green}
\ee
where
$\omega_n=\pi T(2n+1)$ are the Matsubara frequencies ($\hbar=k_B=1$),
$
 \Omega_n=\sqrt{\omega_n^2+\vert\Delta\vert^2},
$
\be
\gamma_n = {\phi\over 2}+{2hd\over {\rm v}_0\cos\varphi}
 - {2i\omega_n d\over {\rm v}_0\cos\varphi}\
\label{gamman}
\ee
\noindent
 and
$\varphi$ is the angle between the direction
$\bf v_0$ and the $x-$axis \cite{zikic,spie96}.

Outside the barrier, $\vert x\vert\geq d$,  we find
\be
g_{\sigma}^{S}(x,\varphi ,\omega_n)  =
g_{\sigma}(\varphi, \omega_n)
e^{\alpha_n(d\pm x)}
+{\omega_n\over\Omega_n}\left(1-e^{\alpha_n(d\pm x)}\right),
\label{greens}
\ee
where  $\alpha_{n} = {2\Omega_{n}/ v_0 \cos \varphi}$ and
$ \sigma=\uparrow, \downarrow \,$ denotes the spin orientations with respect
to the direction of the exchange field.

For the opposite spin direction, the corresponding  Green's
function is obtained by changing $h\to -h$
\be
g_{\uparrow}(h)=g_{\downarrow}(-h). \label{spins}
\ee

\subsection{Densities of states}

The quasiparticle spectrum follows from the retarded Green's functions,
obtained by the analytical continuation of $g_{\downarrow}(h)$ and
$g_{\downarrow}(-h)$. With our definition of $g_{\sigma}$,
the partial  density of states (PDOS), which is the angle-resolved DOS,
is given by
   \be
N_{\sigma}(x, \varphi,  E )=
\lim_{\delta\to 0}\Re g_\sigma (x, \varphi, i\omega_n\to  {E}+i\delta ),
\label{(3.1)}
    \ee
where $E$ is the quasiparticle energy measured from the Fermi level.

DOS is obtained  by averaging PDOS over the angle $\varphi$, assuming
spherical  Fermi surface
           \be
N_{\sigma}(x, {E}) / N_0 = \int_0^{\pi/2} d\varphi
\sin\varphi{\cal D}(\varphi)N_{\sigma}(x, \varphi, {E}),
\label{(3.2)}
           \ee
where $N_0={mk_F / 2\pi^2}$,
and ${\cal D}(\varphi)$ is the normalized barrier
transmission probability \cite{fogel}.
We model the  barrier with a uniform probability distribution ${\cal
D}(\varphi)=1/\int_0^{\varphi_c}\sin\varphi d\varphi$, within an
acceptance cone of angle $2\varphi_c$ about the interface normal, and
zero outside the cone.
The relation between PDOS in $S$:
   \be
N^S_{\sigma}(x, \varphi,  E )=
\lim_{\delta\to 0}\Re g^S_\sigma (x, \varphi, i\omega_n\to  {E}+i\delta ),
    \label{(3.1s)}
    \ee
and in the barrier, Eq. (\ref{(3.1)}), is obtained using Eq.
(\ref{greens}). For $|E| < |\Delta |$ Eq. (\ref{(3.1s)}) reduces to the
simple relation
                  \be
N_{\sigma}^S(x, \varphi, {E}) =e^{\alpha(d\pm x)}
N_{\sigma}(\varphi ,E)      \label {(3.4)}\quad ,
               \ee
where $\alpha = {2{\Omega}/ v_0 \cos\varphi}$, and
${\Omega}=\sqrt {\abs{\Delta}^2-{E}^2} $.
The influence of the barrier is seen only
in the close vicinity of the interface.
Deep in $S$ we obtain, as expected, the well known bulk result
             \be
N^{S}(E) =
{\abs {E}\over {\Omega}}\Theta(\abs {E} -\abs{\Delta})
 \label {(3.5)}.
             \ee

Since PDOS in the electrodes and in the barrier
 are related, we consider in
the following only the position-independent PDOS and DOS in the barrier.

Measuring all energies  in units of $\Delta$,
$\tilde E=E/\Delta$, $\tilde \Omega=\Omega/\Delta$, $\tilde h=h/\Delta$,
 for PDOS we get
               \be
N_{\downarrow}(\varphi,  \abs {\tilde E} <1)=
               {1\over {\tilde \Omega}}\delta({\tilde E}
                +{\tilde \Omega}\cot\gamma )
\label{(3.6)}   ,
               \ee
where $\delta$ is the Dirac delta function, and
               \be
N_{\downarrow}( \varphi, \abs {\tilde E} >1)=
            {-i\abs {\tilde E} \tilde \Omega\over {\tilde E}^2-\cos^2\gamma}  , \label{(3.7)}
               \ee
where
$\gamma =\phi /2 -\tilde d({\tilde E} -\tilde h)/\cos\varphi$,  $\tilde d =2d\Delta/v_0=
(2/\pi)d/\xi_0$, and $\xi_0=v_0/\pi\Delta$.

Bound states exist only for energies $|\tilde E|<1$   and satisfy the equation
            \be
\tan\gamma =-{{\tilde \Omega}\over\tilde  {E}}\quad      ,      \label{(3.8)}
            \ee
corresponding to the poles in the analytical
 continuation of the Green's function,  Eq.(\ref{green}).
 A more convenient form of {E}q. (\ref{(3.8)}) is
         \be
\sin\gamma =\pm {\tilde \Omega}, \quad \cos\gamma =\mp \tilde {E}   \label{(3.9)}
          \ee
with the condition
          \be
{\hbox{sign}(\tilde {E})}\sin\gamma\cos\gamma \le 0\quad .
       \label{(3.10)}
          \ee
Note that the bound states with opposite spin orientation, spin up, are
given by {E}q. (\ref{(3.8)})  (or by {E}qs. (\ref{(3.9)}), (\ref{(3.10)}))
with $h\to -h$. There is a spin splitting of bound
states.

Another consequence of {E}q. (\ref {(3.9)}) is that
bound states at the Fermi level (ZES) appear
for $\cos \gamma=0$. For $\varphi=0$ this gives
 \be
{\phi\over 2}\pm\tilde d \tilde h =(2k+1){\pi \over 2},\qquad
 k=0,\pm 1, \pm 2,...\quad .
\label{(3.11)}
   \ee

Calculating DOS for  $\varphi_c=\pi/2$, we find
         \be
&&N_{\downarrow}(\abs {\tilde E} <1) / N_0 =  \label{(aa)} \\
 & = & 2\pi\tilde d\abs{{\tilde E}-\tilde h}
 \sum_{n=-\infty}^\infty{\Theta(A)
 \over \left({\phi/ 2} + n\pi +\arctan{{\tilde \Omega}/ {\tilde E}}\right)^2  }
 \nonumber
         \ee
where  $A = \hbox{sign}({\tilde E} - \tilde h)\left(\Phi +n\pi \right)$
and $\Phi=\arctan {\Omega} /{\tilde E} + \phi /2 -\tilde d({\tilde E} -\tilde h)$.
Let us  denote  by ${\tilde E}_i$ the zeros of {E}q. (\ref{(3.8)}) for $\varphi =0$,
which can be written in the form $\Phi({\tilde E} _i)+n_i\pi =0,\, n_i
=0,\pm 1,...$. Then,
{E}q. (\ref{(aa)}) can be rewritten (see Appendix A) in the form
          \be
N_{\downarrow}(\abs {\tilde E} <1) / N_0  =  2\pi\tilde d\abs{{\tilde E}-
\tilde h}  \left\{
\sum_{i=1}^m{\Theta({\tilde E} -\tilde h)\Theta({\tilde E}_i - {\tilde E}) +
\Theta(\tilde h-{\tilde E})\Theta({\tilde E} -{\tilde E}_i)
\over \left({\phi/ 2} + n_i\pi
+\arctan{\tilde \Omega/{\tilde E}}\right)^2} +\right. \label{(3.13)}\\
 + \left. \sum_{n=-\infty}^{n^-}{\Theta (\tilde h-{\tilde E}) \over \left(
{\phi/ 2} + n\pi+\arctan{\tilde \Omega/{\tilde E}}\right)^2}
 + \sum_{n=n^+}^\infty {\Theta ({\tilde E} -\tilde h) \over
\left({\phi/ 2} + n\pi +\arctan{\tilde \Omega/{\tilde E}}\right)^2}\right\},
\nonumber
              \ee
where $m$ is the total number of bound states ${\tilde E}_i$,
and  integers $n^+$, $n^-$  are defined by the conditions
$\Phi+n\pi\geq 0$ for $n\geq n^+$  and
$\Phi+n\pi\leq 0$ for $n\leq n^-$, respectively.
Integers $n_i\in [n^-+1, n^+-1]$ are chosen so that
 $\sgn {\Phi({\tilde E}) +n_i\pi} = \sgn {{\tilde E} -\tilde h}$.

The physical meaning of this result is the following: due to the
isotropy of the order parameter the averaging over the quasiparticle
propagation angle $\varphi$ results in an average over one-chanel weak
links of different lengths. Therefore there is no  qualitative
difference between DOS and PDOS, the peaks in DOS corresponding to peaks
in PDOS for $\varphi=0$ situated at the energies $\tilde E_i$.

For normal metal barrier, $\tilde h=0$, our result agrees with Refs.
[24]  and [22] with small corrections of DOS for ${\tilde
E}\geq{\tilde E}_i$. For $\phi=0$     the  total number $m$  of bound
states  depends on the thickness, $m=2([{\tilde d /\pi}]+1)$,  where the
symbol $[A]$ represents the greatest integer less than $A$ \cite{arnold}.
In this case, ZES may appear only when the phase difference $\phi$ at
the link is different from zero, {E}q. (\ref{(3.11)}).

For ferromagnetic metal barrier, $\tilde h\neq 0$, the ground state
phase difference $\phi_{gs}$ has been calculated \cite{spie96}  by
minimizing the energy of the link, and it was found that $\phi_{gs}=0$
for weak barrier influence, $2\tilde h\tilde d<1$, and $\phi_{gs}=\pi$ for
stronger influence, $1<2\tilde h\tilde d<4$. Larger values of $2\tilde
h\tilde d$ would correspond to the decoupling of $S$ electrodes \cite
{10}.
Combining this  result with the condition for ZES, {E}q. (\ref{(3.11)}),
one can easily see that ZES can occur for $\phi\neq \phi_{gs}$ only. Due
to the spin splitting, SPGS may appear \cite{10,11,dwa}. Analyzing {E}q.
(\ref{(3.8)}) for $\varphi=0$, we find that for $\tilde d\geq{\pi / 2}$
there is always at least one bound state for each spin orientation in
the interval $-1<\tilde E<0$. Therefore,  the spin polarized ground
state, where only one spin orientation negative energy levels are
populated, may exist only in thin barriers, $\tilde d< {\pi / 2}$. They
actually appear (see Appendix B), with spin up orientation, provided
that ${\phi/2}+\tilde d\tilde h$ belongs to the (open) interval $\left(
(2k+1)\pi/2, (k+1)\pi -\tilde d\right)$, whereas ${\phi/2}-\tilde
d\tilde h$ lies outside the above interval, and {\it vice versa} for
spin down orientation.

We emphasize that all above analytical results are in complete agreement
with numerical calculations of PDOS and DOS, which we performed directly
from  {E}qs. (\ref{green}) and (\ref{(3.1)}). In numerical calculations,
we have everywhere taken $\varphi_c=\pi /2$. For smaller $\varphi_c$
there is no qualitative difference in DOS , the number and the positions
of the peaks are the same, the  relevant direction being that
perpendicular to the barrier, $\varphi =0$. This is not the case for
$d-$wave pairing in cuprates, as shown in Ref. \cite{zikic}.

\subsection{Supercurrents}

The supercurrent trough the $S/F/S$ junction is also obtained from
the  Green's functions  in the barrier, Eqs. (\ref{green}) and
(\ref{spins}).
$$
{\bf j} = -2ie\pi N_0T\sum_{\omega_n}
\left\langle {\bf v}_0 {g_\uparrow
+g_\downarrow\over2}\right\rangle ,
$$
where $\langle\cdots \rangle$  denotes the angular averaging over
the Fermi surface.
Putting $ u=1/\cos\varphi$ we get
\be
I = {{\pi I_0 T}\over {4 \Delta}}\sum_{\omega_n}\sum_{\sigma}
 \int^\infty_1 {\Im [g_{\sigma}(u)
+g_{\sigma}(-u)]\over
u^3 }du, \quad \sigma=\downarrow, \uparrow .
\label{supercurrent}
\ee

This gives the supercurrent through the barrier of the area $S$, $I=jS$
as a function of $\phi$,  temperature $T$, and of parameters measuring
the influence of the F barrier, $h$ and $d$. The normalizing
current is $I_0=2\Delta/e R_N$, and the normal resistance is given by
$R_N^{-1}=e^2{\rm v}_0N_0S$, where $N_0$ is the density of states at
the Fermi surface.

From {E}q. (\ref{supercurrent}) we see that poles of
$g_{\sigma}$ determine both the quasiparticle energy spectrum and
the phase and exchange field dependence of the supercurrent.
That the supercurrent is carried by the Andreev bound state is particularly
clearly seen at $T=0$.
Taking $T\sum_{\omega_n} \to \int d\omega /2\pi$, it is easy to
show that Eq. (\ref{supercurrent}) at zero temperature can be rewritten
in the form
        \be
I(\phi)={{\pi \over 2}}I_0\sum_{j}\int^\infty_1 {{du\over u^3}
\left[ {\partial \tilde E_{j,\downarrow}\over \partial \phi}+
{\partial \tilde E_{j,\uparrow}\over \partial \phi}\right] },
\label{andrcurrent}
         \ee
where $\tilde E_{j,\downarrow}$ and $\tilde E_{j,\uparrow}$  are negative energy levels, obtained
from  Eq. (\ref{(3.9)}) with $\gamma=\gamma(\pm h)$. Since the pair
potential is assumed to be spatially constant
in the $S$ electrodes, the free energy of the junction is related to the
supercurrent by the well-known relation
            \be
I(\phi)=2e{\partial F\over \partial \phi}
\label{partial}
       \ee and
        \be
2e F(\phi)={\pi \over
2}I_0\sum_{j}\int^\infty_1 {{du\over u^3}[\tilde E_{j,\uparrow}
 +\tilde E_{j,\downarrow}]}.
\label{energye}
            \ee
In particular, in the limit of strong exchange field, $\tilde h\gg 1$,
we obtain
               \be
I={\pi\over 4}I_0\int^\infty_1 {{du\over u^3} \left[
 \sin({{\phi\over 2}+ \tilde d \tilde hu}) \sgn{\cos({{\phi\over 2}+
 \tilde d \tilde hu})}
 +\sin({{\phi\over 2}- \tilde d \tilde hu})\sgn{\cos({{\phi\over 2}- \tilde d
 \tilde hu})}\right] }
 \label{zlarge}
              \ee
and
           \be
2e F(\phi)=-{\pi\over 2}I_0\int^\infty_1 {{du\over u^3}\left[
\left|\cos({{\phi\over 2}+\tilde d \tilde hu})\right| +
\left|\cos({{\phi\over 2}-\tilde d \tilde hu})\right|\right]}.
\label{energy}
\ee
Therefore,
even in this limit, the supercurrent is carried
by the Andreev bound states of both spin orientations, and the phase
dependence is not sinusoidal, in contrast to the results of Tanaka and
Kashiwaya for ferromagnetic tunnel junctions \cite{11}.

\section{Summary and discussion}

In this Section, we discuss the results obtained for $S/N/S$ and
$S/F/S$  weak links, with  thin metallic barrier, $2d/\xi_0<\pi$
($\tilde d<1$).

The results of numerical calculation of the supercurrent at $T=0$
as a function of $\phi$, for some characteristic values of $h$, are shown
in Fig. 2, and the results for PDOS and DOS in Figs. 3-6. In all
examples we have taken $2d=\xi_0$ ($\tilde d=1/\pi$).

In both cases of normal ($h=0$) or ferromagnetic $(h\neq 0)$ metal
barrier, the quasiparticle  spectrum in the barrier is gapless, with
characteristic peaks inside the superconducting gap,
$|E|<\Delta$. Due to the isotropy of the order parameter in $S$, there
is no qualitative difference between PDOS for $\varphi=0$ and DOS. The
peaks in PDOS, corresponding to the bound states, are reflected in DOS
as peaks placed at the same energies. This is illustrated for $h=0$ in
Fig. 3, but holds for $h\neq 0$ as well.

For thin normal metal barrier,  the number  of bound states is $m=2$ or
$m=1$. The spectra are strongly influenced by the phase difference
$\phi$. A single peak, corresponding to ZES,
appears for $\phi=\pi$ at the maximum  of the supercurrent through the link,
Fig. 2(a).  In the ground state, $\phi=\phi _{gs}=0$, DOS starts
from zero at ${E}=0$ and has two peaks at ${E} \neq 0$, Fig. 3.

For ferromagnetic metal barrier, $h \neq 0$, the bound states for two
spin orientations (spin up and spin down with respect to the exchange
field orientation) are no  more degenerate. However, both kinds of bound
states contribute to the supercurrent through the link. The dependence
$I(\phi)$ is shown in Fig. 2(b) for $\phi_{gs}=0$ and for $\phi_{gs}=\pi$.
DOS  curves are splitted with respect to the case $h=0$, and the
number of peaks is doubled. An example with $\phi_{gs}=0$ is
shown in Fig. 4.

Now, the conditions for ZES, Eq. (\ref{(3.11)}), are $\phi\pm 2\tilde
d\tilde{h}= \pi$.
The main effect of $h \neq 0$ is to induce an
additional, "magnetic" phase difference $\phi \to \phi \pm 2\tilde d
\tilde h$,
for the electrons traveling in the direction $\varphi =0$. Thus, the
formation of ZES is due to an appropriate value of the total phase
difference. When a  ZES appears for a given value of $\phi$ and $\tilde h$ for
one spin direction, there is no ZES for the same $\phi$ for the opposite
spin direction, $h\to -h$, except if $\phi=0$ and $\tilde d \tilde h=\pi/2$.
The latter case is illustrated in Fig. 5  for $\tilde h=\pi^2/2$.
Here $\phi_{gs}=\pi$, and ZES is obtained outside the ground state, for
$\phi=0$.

Appearence of spin polarized states, with a single spin orientation at
negative energies, is illustrated in Fig. 6, for $\tilde h=1.5$. For given
$\tilde h$
and $\tilde d$, the spin orientation of the polarized  state strongly
depends on the phase difference $\phi$. In the given example, the
negative energy state  in the ground state, $\phi=\phi_{gs}=0$, has the
spin down orientation, and for $\phi=\pi$, the spin up orientation, in
accordance with the condition (\ref{B.2}) in the Appendix B.

In conclusion, we have analyzed the quasiparticle spectra in clean
$S/F/S$ contacts, with $s-$wave pairing in $S$ electrodes. In this  case
DOS for each spin orientation is asymmetrical, and strongly exchange
field and    phase  dependent. For both ferromagnetic and normal metal
barriers we have found  conditions for the formation of zero energy
bound states, and in the case of ferromagnetic barrier, for the
formation of spin polarized ground states. In all cases, zero energy
bound states  appear only outside  the weak link ground state.

\section*{Appendix A  }

In this Appendix we show how in the case of $s-$wave pairing, DOS for
$\tilde {E} <1$, {E}q. (\ref{(aa)}), can be expressed  in terms of bound
states energies, {E}q. (\ref{(3.13)}). Since the function
$\Phi=\arctan{\tilde \Omega} /\tilde {E} + \phi /2 -\tilde d(\tilde {E}
-\tilde h)$ is bounded in
the interval $|{\tilde E} |<1$, there is a minimum integer $n^+$ such that
$\Phi _{\rm min} +n^+\pi\ge 0$. This means that for each $n\ge n^+$ and
each ${\tilde E}$, $\Phi +n\pi\ge 0$. Similarly, there is a maximum integer
$n^-$ such that for each $n\le n^-$, and for each ${\tilde E}$, $\Phi +n\pi\le
0$. Thus, {E}q. (\ref{(aa)}) can be written in the form
        \be
N_{\downarrow}  (\abs{\tilde E} <1) / N_0  = 2\pi\tilde d\abs{{\tilde E}-
\tilde h}
 \left\{ \sum_{n=-\infty}^{n^-} {\Theta
(\tilde h-{\tilde E})\over \left( {\phi/ 2} +
n\pi+\arctan{\tilde \Omega/{\tilde E}}\right)^2} + \right.\nonumber\\
+\left.
\sum_{n=n^+}^\infty {\Theta ({\tilde E} -\tilde h) \over \left({\phi/ 2} + n\pi
+\arctan{\tilde \Omega/{\tilde E}}\right)^2} +
\sum_{n=n^- + 1}^{n^+ -1}{\Theta (A) \over \left({\phi/ 2} + n\pi
+\arctan{\tilde \Omega/{\tilde E}}\right)^2}
\right\},\label{(4.4c)}
         \ee
    where
$A=\hbox{sign}({\tilde E} - \tilde h)\left(\Phi  +n\pi \right)$. In this expression
the first two sums are continuous functions in the whole interval $|{\tilde E}|<1$.
In the third sum, all  terms for which the condition $\sgn{\Phi(\tilde E) +n\pi}
= \sgn{\tilde E -\tilde h}$ is not satisfied are missing, giving rise to finite
discontinuities. These discontinuities occur at bound states
energies ${\tilde E}={\tilde E}_i$ for $\varphi=0$, as can be seen by
considering the behavior of the
function $\Phi(\tilde E)$, which is monotonous and descending in the intervals
$(-1,0)$ and $(0,1)$, with a discontinuity at ${\tilde E} =0$. In fact,
$\Phi(\tilde E) +n_i\pi$ changes sign in the vicinity of ${\tilde E}_i$,
since by definition $n_i$ is an integer such that $\Phi ({\tilde
E}_i)+n_i\pi =0$. For $\Phi
({\tilde E})+n_i\pi<0$ and ${\tilde E}-\tilde h>0$, as well as for
$\Phi({\tilde E})+n_i\pi>0$ and
${\tilde E}-\tilde h<0$, the $n_i^{\rm th}$ term in the third sum in
{E}q. (\ref{(4.4c)}) is missing due to the step function $\Theta (A)$.
Therefore, the sum is discontinuous at ${\tilde E} ={\tilde E}_i$,
and the third sum over $n$ in {E}q. (\ref{(4.4c)}) can be written as a sum
over ${\tilde E}_i$ in {E}q. (\ref{(3.13)}).

\section*{Appendix B  }

In this Appendix we derive the conditions for the appearence of the
spin polarized states from {E}q. (\ref{(3.8)}), written in the form
$\Phi (\tilde E)=n\pi$, $n=0$, $\pm 1$, $\pm 2$,...,
where
$\Phi (\tilde E)=\arctan{{\tilde \Omega} /{\tilde E}} +
 \phi /2 -{\tilde d}({\tilde E} \mp {\tilde h})$ for spin
down (up) orientation. Let us denote the greatest number $n$ for which
$\Phi (\tilde E=-1)> n\pi$ by $n_0=\left[ {\Phi (\tilde E=-1)/\pi} \right ]$,
where the
symbol $[A]$ represents the greatest integer less than $A$.

Since $\Phi (\tilde E)$ is monotonously decreasing in the energy intervals
$(-1,0)$ and $(0,1)$ (with a discontinuity at $\tilde E=0$), there would be no
solution in the negative energies interval,
if $\Phi (\tilde E=0^-)>n_0\pi$, or explicitly if
  \be
{{\phi/2\pm \tilde d\tilde h-\pi}\over \pi}>\left [{{\phi/2\pm \tilde d
\tilde h-\tilde d}\over \pi}\right ].
\label{B.1}
  \ee
One can check that the condition (\ref{B.1}), which means that there
is no spin down (up) bound states in the negative energies interval $(-1,0)$,
is satisfied provided that
  \be
\phi/2\pm \tilde d\tilde h \in \left( (2k+1)\pi/2, (k+1)\pi-\tilde
d\right).
\label{B.2}
  \ee
Therefore, to have a spin up polarized ground state, $\phi/2+ \tilde d\tilde h$
must be inside and $\phi/2- \tilde d\tilde h$ outside the above (open)
interval, and {\it vice versa} for the spin down polarized ground state.

Similar conditions hold in the positive energies interval $(0,1)$. There
is no spin down (up)  states provided that
  \be
{{\phi/2\pm \tilde d\tilde h-\tilde d}\over \pi}>\left [{{\phi/2\pm \tilde d
\tilde h+\pi/2}\over \pi}\right ],
\label{B.3}
  \ee
i. e.
  \be
\phi/2\pm \tilde d\tilde h \in \left( k\pi+\tilde d, (2k+1)\pi/2\right).
\label{B.4}
  \ee

For $\tilde d>\pi/2$ neither of the conditions (\ref{B.2}) nor
(\ref{B.4}) is
satisfied, so  there is no spin polarized states. Note that the condition
for ZES is obtained when $\phi/2\pm \tilde d\tilde h =(2k+1)\pi/2$, i. e. at
the  only common point of two intervals in (\ref{B.2}) and (\ref{B.4}).

\eject

\begin{figure}
    \caption{ \label{sl 1} }
Schematic illustration of the weak link.
\end{figure}

\begin{figure}
   \caption{ \label{sl 2} }
Normalized supercurrent at zero temperatures for (a) $S/N/S$, and (b)
$S/F/S$  weak links.
In both cases the reduced thickness of the barrier is $2d/\xi_0 = 1$.
For $S/F/S$ two examples are presented:
a "0-contact", $\phi_{gs}=0$, with $h/\Delta =1.5$ (dotted curve),
and a "$\pi$-contact", $\phi_{gs}=\pi$, with $h/\Delta=4.7$ (solid curve).
\end{figure}

\begin{figure}
 \caption{ \label{sl 3} }
Densities of states in the barrier of $S/N/S$ weak link with $2d/\xi_0 =1$:
(a) PDOS for $\varphi=0$,
(b) DOS averaged over the spherical Fermi surface for
$\phi=0$ (solid curves), and for $\phi=\pi$ (dashed curves).
\end{figure}

\begin{figure}
   \caption{ \label{sl 4}         }
DOS in the ferromagnetic barrier of $S/F/S$ weak link with
$2d/\xi_0 = 1$ and $h/\Delta =0.6$:
(a) ground state, $\phi =0$,
(b) $\phi =\pi$. Spin splitting: $\sigma=\uparrow$ (solid curves) and
$\sigma=\downarrow$ (dotted curves).
\end{figure}

\begin{figure}
   \caption{ \label{sl 5}  }
DOS in the ferromagnetic barrier of $S/F/S$ weak link with
$2d/\xi_0 = 1$ and $h/\Delta=\pi^2/2$: (a) $\phi=0$, (b) ground
state, $\phi =\pi$. Spin splitting: $\sigma=\uparrow$ (solid curves) and
$\sigma=\downarrow$ (dotted curves).
Note that for $\phi=0$ ZES appear for both spin orientations in
accordance with Eq. (\ref{(3.11)})
\end{figure}

\begin{figure}
   \caption{ \label{sl 6}  }
DOS in the ferromagnetic  barrier of $S/F/S$ weak link with
$2d/\xi_0 = 1$ and $h/\Delta =1.5$: (a) ground state, $\phi=0$,
(b) $\phi=\pi$. Spin splitting: $\sigma=\uparrow$ (solid curves)
and $\sigma=\downarrow$ (dotted curves).
\end{figure}

\end{document}